# Kirigami Metamaterials for Reconfigurable Toroidal Circular Dichroism

Liqiao Jing[1#], Zuojia Wang[2#*], Bin Zheng[1,3*], Huaping Wang[1,4], Yihao Yang[1], Lian Shen[1], Wenyan Yin[1], Erping Li[1], Hongsheng Chen[1,3*]

[1]Key Lab. of Advanced Micro/Nano Electronic Devices & Smart Systems of Zhejiang, Zhejiang University, Hangzhou 310027, China

[2]School of Information Science and Engineering, Shandong University, Jinan 250100, China

[3]State Key Laboratory of Modern Optical Instrumentations, The Electromagnetics Academy at Zhejiang University, Zhejiang University, Hangzhou 310027, China

[4]Institute of Marine Electronics Engineering, Zhejiang University, Hangzhou 310058, China

[#] Co-first authors.
[*]Corresponding authors:

Name: Zuojia Wang
Full postal address: 27 Shanda South Road, Jinan 250100, China
Telephone: +86- 13616518084
E-mail address: z.wang@sdu.edu.cn

Name: Bin Zheng
Full postal address: 38 Zheda Road, Hangzhou 310027, China
Telephone: +86-15868184212
E-mail address: zhengbin@zju.edu.cn

Name: Hongsheng Chen
Full postal address: 38 Zheda Road, Hangzhou 310027, China
Telephone: +86-571-87951013
E-mail address: hansomchen@zju.edu.cn

**ABSTRACT**

The ancient paper craft of kirigami has recently emerged as a potential tool for the design of functional materials. Inspired by the kirigami concept, we propose a class of kirigami-based metamaterials whose electromagnetic functionalities can be switched between nonchiral and chiral states by stretching the predesigned split-ring resonator array. Single-band, dual-band and broadband circular polarizers with reconfigurable performance are experimentally demonstrated with maximum circular dichroisms of 0.88, 0.94 and 0.92, respectively. The underlying mechanism is explained and calculated via detailed analysis of the excited multipoles, including the electric, magnetic, and toroidal dipoles and quadrupole. Our approach enables tailoring the electromagnetic functionalities in kirigami patterns and provides an alternate avenue for reconfigurable optical metadevices with exceptional mechanical properties.

**KEYWORDS:** kirigami, metamaterials, chiroptical response, reconfigurable

**INTRODUCTION**

Metamaterials are artificial materials engineered at the subwavelength scale to achieve electromagnetic functionalities[1]. Several novel optical phenomena have been observed in metamaterial-based devices, such as negative refraction[2-4], superlens[5], invisibility cloaking[6-8], and strong chiroptical responses[9,10]. Although bulk metamaterials show intriguing optical properties, the complexity of fabrication and large loss in metallic meta-atoms hamper their applications in practice. Recently, new degrees of freedom have been attained by introducing abrupt phase discontinuities on metasurfaces, the two-dimensional counterparts of metamaterials. Metasurfaces show great capabilities in wavefront manipulation and reduce the complexity of fabrication[11-15]. Spin-selective absorption has been demonstrated by designing chiral meta-atoms, with promising potential applications in polarimetry, circular polarization detectors and chiral cavities[16-19]. However, structural modification is generally challenging once metamaterials or metasurfaces are fabricated, rendering them functionally nonreconfigurable.

Reconfigurable metamaterials are designed to achieve dynamic control over the physical properties to realize multiple functions in one metadevice[20]. Tuning methods include the use of capacitors[21], semiconductors[22], phase-change materials[23,24] and ferromagnetic/ferroelectric materials[25]. However, compared with the surrounding media and metamaterial constituents, most of these methods suffer from a limited tuning range because the variation is usually very small. Another strategy is changing the structural shapes for reconfigurable functionality. Recently, origami provided an alternative approach to construct strong, lightweight, and tunable three-dimensional (3D) blocks from flat sheets[26,27]. By applying prescribed sequences of folds to flat surfaces, researchers demonstrated flexible and efficient control over mechanical[26,28], electronic[29], acoustic[30], superconducting[31-33] and electromagnetic functionalities[34]. Although the design capacity of origami is remarkable, achieving complex target shapes with only folds is mathematically challenging. Such complex folding patterns require a convoluted series of deformations from the flat to folded shapes, making fabrication difficult. Different from origami, kirigami is an art form that introduces cuts into folding processes, providing extra degrees of freedom in 3D shape construction[35]. It allows for similarly complicated shapes to be formed with greatly reduced complexity in the design process and less wasted material[36,37]. When applying sufficiently large amounts of stretching, buckling is triggered, resulting in the formation of a 3D structure comprising a well-organized pattern of mountains and valleys. Kirigami is a highly promising technique to design complex 3D metadevices with reconfigurable functionalities but has not been applied in the design of electromagnetic metamaterials with extraordinary chiral properties, especially for reconfigurable toroidal circular dichroism.

A chiral structure can be modeled as electric and magnetic dipoles with parallel or antiparallel orientations of comparable magnitude. Different absorption occurs when circularly polarized waves pass through such a structure. This is the widely discussed phenomenon of conventional circular dichroism. However, for toroidal circular dichroism, the chiroptical effects are mostly attributed to the combination of toroidal dipoles and other higher order electric multipoles.

In this letter, we propose and demonstrate a general class of kirigami-based chiral metamaterials (KCMMs) whose electromagnetic performance can be switched between nonchiral and chiral states at single-band, dual-band and broadband wavelengths. Moreover, for the first time, we propose and investigate the reconfigurable toroidal dipole moments based on the kirigami. Split-ring resonators (SRRs) are periodically arranged on thin, foldable sheets. When transforming the 2D metasurface to 3D kirigami-based metamaterials, the resonant modes exhibit a gradually enhanced chiroptical response. The underlying mechanism is explained via detailed analyses of the excited multipoles, including the electric, magnetic, and toroidal dipoles and electric quadrupole. The nonradiating feature and high flexibility of the toroidal geometry will be useful in many applications, such as lasers[38,39], ultrasensitive biosensors and nonlinear effects[40,41]. The reconfigurable circular dichroism generated by toroidal dipoles offers an alternative approach to broadband chiroptical responses beyond the widely adopted methods based on parallel electric and magnetic dipoles. Circular dichroisms of 0.88, 0.94 and 0.92 have been experimentally observed for single-band, dual-band and broadband configurations, respectively.

**MATERIALS AND METHODS**

The schematic of the KCMMs is illustrated in Figure 1. The functionality of the proposed KCMMs is to totally transmit the designated circularly polarized wave and reflect the other spin state with maximum efficiency. Split-ring resonators are adopted as the basic meta-atoms printed on a thin and flexible dielectric substrate. Before cutting and folding, the metasurface is achiral because of its mirror symmetry with respect to the *yz* plane. Subsequently, the 2D metasurface is transformed into 3D geometries by introducing cuts at the boundary between neighboring meta-atoms in the *y* direction. The kirigami approach is considerably diverse; we investigate three types of KCMMs that are cut and folded from 2D metasurfaces. Type-I KCMMs represent kirigami structures whose neighboring units are connected by the midpoints of the sides at the cut boundary, whereas type-II KCMMs are based on buckling-induced kirigami in which the meta-atoms are connected by the vertices of the squares. Foldable 3D structures are formed by stretching the cut materials. Both type-I and type-II KCMMs contain four SRRs in a unit cell. To obtain toroidal dipole responses, type-III KCMMs are formed by stacking two type-I layers. The mirror symmetry is broken by introducing cuts and folds along different axes. Notably, all KCMMs can be folded into two kinds of chiral enantiomers, which are mirror images of each other. For simplicity, we label the two enantiomers as R-handed and L-handed. Additional details on the geometric dimensions of KCMMs and the folding process can be found in the Supporting Information (Figure S1, S3).

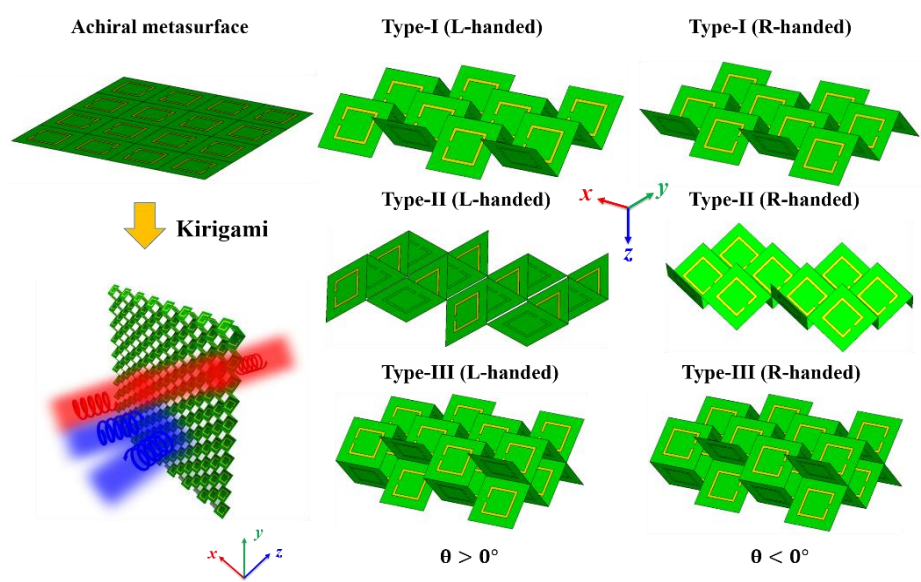

**Figure 1.** Schematic of the construction of KCMMs. Split-ring resonators are periodically arranged to form an achiral metasurface (left top panel). The 2D metasurface can be continuously transformed into 3D chiral metamaterials: type-I KCMM (top panel), type-II KCMM (middle panel) and type-III KCMM (bottom panel). Chirality switching is achieved by changing the deformation direction of the planar metasurface.

Proof-of-concept KCMMs were designed in the microwave region. The unit cell of the 2D metasurface is composed of four SRRs with side length l and width w. The gap size of each SRR is $g = 1$ mm, and the lattice constant is 20 mm. This unfolded metasurface has mirror symmetry around the *yz* plane and hence does not exhibit intrinsic chirality. Full-wave numerical simulations were conducted using the commercial software CST Microwave Studio. In simulation, SRRs are assumed to be standing in free space, and the influence of the ultrathin dielectric substrate is neglected. This is reasonable because the dielectric substrate only influences the gap capacity of the SRR and thus produces only a slight shift of the resonant frequency.

**RESULTS**

The simulation results for the three types of KCMM are illustrated in Figure 2. In the folded state of $\theta = 45°$, the type-I KCMMs exhibit chiroptical responses at the resonant frequency of 6.78 GHz, with opposite handedness in the two enantiomers. For the L-handed enantiomer, as shown in Figure 2(a), there is a left-handed resonant mode in which right-handed circularly polarized (RCP) waves are highly transmitted ($t_{RR}$=0.95) and left-handed circularly polarized (LCP) waves are perfectly reflected ($t_{LL}$=0.03). Chirality switching into the R-handed enantiomer is realized by changing the folding direction, resulting in the completely reversed chiroptical response, as shown in Figure 2(b). Both the type-I KCMM enantiomers demonstrate only one strong chiral resonant mode. In the type-II KCMMs, buckling is introduced to induce out-of-plane rotations of both the square domains and the cuts. As shown in Figure 2(e), a second chiral resonant mode occurs at a higher frequency of 7.8 GHz with opposite handedness. When linearly polarized waves impinge on the L-handed enantiomer of the type-II KCMM, LCP and RCP waves are transmitted at 6.78 GHz and 7.8 GHz, respectively. Similarly, the R-handed enantiomer modulates waves with opposite handedness (Figure 2(f)). Therefore, type-II KCMMs can act as dual-band bifunctional circular polarizers that filter different spin states at two resonant frequencies. Although type-I and type-II KCMMs show the inspiring ability to manipulate chiroptical effects, the assembly of multilayer KCMMs could offer striking bandwidth enhancement. The type-III KCMMs are stacked from two type-I KCMMs with identical handedness. Two resonant modes of the L-handed type-III occur at 6.40 and 7.12 GHz (Figure 2(i)). Unlike the type-II KCMMs, both modes show the left-handed feature and hence could be combined to achieve a broadband feature. In contrast, stacking two R-handed type-I KCMMs can produce an R-handed type-III KCMM that behaves in an opposite manner (Figure 2(j)).

To illustrate the abilities of the KCMMs as circular polarizers, we investigate the circular dichroism (CD) spectra, which are calculated by CD = $|t_{RR}|^2$ - $|t_{LL}|^2$. The dependence of performance on the folding angle is plotted in Figure 2 (c, g, h). It demonstrates that the chiroptical effects of R-handed and L-handed enantiomers are complementary to each other. The chiral responses were gradually enhanced as the folding angle increased and reached their maximum at approximately 70 degrees. For the type-I and type-II KCMMs, the resonant frequencies gradually shifted as the folding angle increased, indicating that the operating frequencies of the single-band and dual-band circular polarizers can be flexibly tuned by adjusting the deformation states. The broadband performance of the type-III KCMMs is also well preserved at different folding states. The CD curves for three states (θ = 0°, 45°, -45°) of the KCMMs are plotted in Figure 2(d, h, l). When θ = 45°, the maximum CDs for type-I, type-II and type-III KCMMs are 0.90, 0.89 and 0.94, respectively. Such prominent performance makes KCMMs good candidates for reconfigurable circular polarizers. Animations can be found in the supplementary materials (video 1, 2, 3).

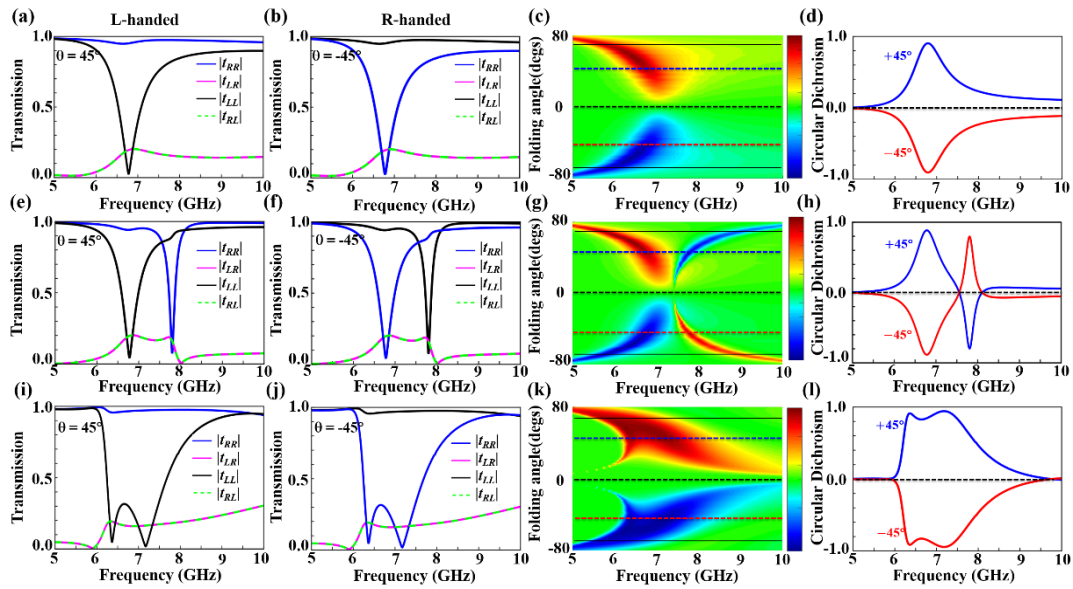

**Figure 2.** Simulated chiral responses of the kirigami-based metamaterials. (a-d) Type-I KCMMs. (e-h) Type-II KCMMs. (i-l) Type-III KCMMs. (a, e, i) the transmission spectra of the L-handed enantiomers and (b, f, j) R-handed enantiomers. The dependence of the CD spectra on the folding angle is shown in (c, g, k), and CD curves for selected angles are shown in (d, h, l).

To demonstrate the chiroptical responses experimentally, samples were fabricated based on printed circuit board (PCB) technology. Copper split-ring resonators (thickness 0.035 mm) were periodically printed on a type of polyimide film (thickness 0.05 mm) with a permittivity of 3.5. Measurements of chiroptical responses and chirality switching of the KCMMs were implemented in a microwave chamber with a vector network analyzer. KCMMs consisting of 10×20 unit cells were designed along the x- and y-axes, respectively. Additional experimental details are described in the Supporting Information. The results are illustrated in Figure 3. As expected, when the 2D metasurface is deformed to 3D geometries, the mirror symmetry of the structure is broken, and the chiroptical response emerges. At a resonant frequency of 6.4 GHz, the L-handed enantiomer of the type-I KCMM blocks most of the LCP waves, whereas the R-handed enantiomer is more opaque for the other spin state (Figure 3(a, b)). The measured CD spectra plotted in Figure 3(c) demonstrate the performance of chirality switching at the resonant frequency, as predicted by the preceding simulations (Figure 2d). For the L-handed type-II KCMM enantiomer, the first resonant frequency is also at approximately 6.4 GHz, where most of the RCP waves are transmitted. The situation is reversed for the second resonant mode at approximately 7.4 GHz, where the type-II KCMM reverses its chirality handedness (Figure 3(d, e, f)). For the type-III KCMMs, broadband chiroptical responses occur between two resonant frequencies of 6.2 GHz and 6.7 GHz. As expected, the L-handed and R-handed enantiomers behave as broadband circular polarizers that transmit only RCP and LCP waves (Figure 3(g, h)), respectively. The measured CD spectra are plotted in Figure 3(i) and are highly consistent with the theoretical prediction.

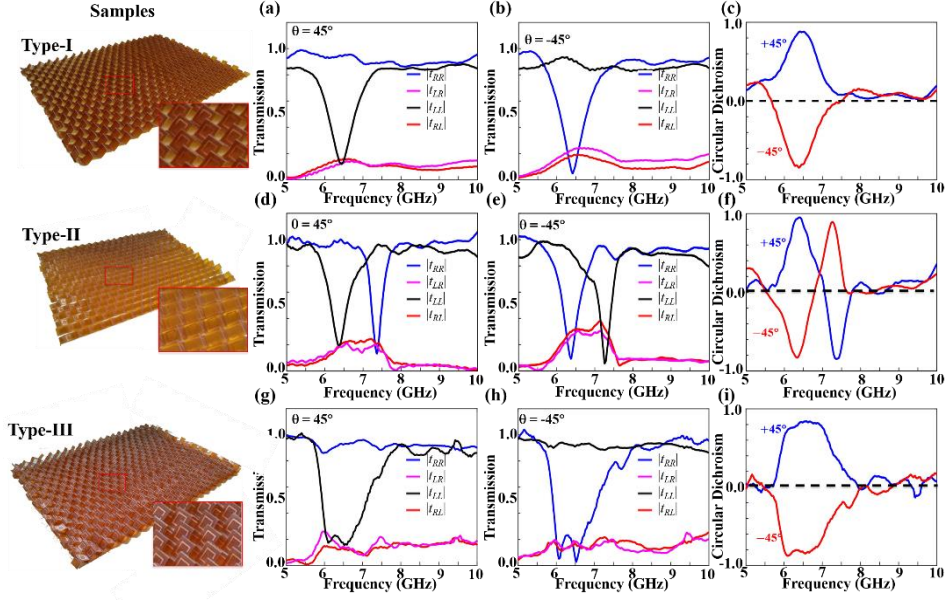

**Figure 3.** Photographs of the fabricated KCMMs and experimental demonstration of the reconfigurable chiroptical responses in kirigami-based metamaterials. (a-c) Type-I KCMMs. (d-f) Type-II KCMMs. (g-i) Type-III KCMMs. (a, d, g) show measured transmission spectra of the L-handed enantiomers, and (b, e, h) correspond to the R-handed enantiomers. (c, f, i) show measured CD spectra, which were calculated from the measured transmission spectra.

Split-ring resonators produce effective electric, magnetic and toroidal dipoles at the resonance. To determine the origin of the circular dichroism at these resonances in our KCMMs, we conducted a dipolar analysis to determine the underlying mechanism. The specific combination of the electric and magnetic dipoles, that is, parallel or antiparallel with comparable magnitudes, can induce strong chiral responses similar to those in natural chiral molecules[42,43]. For simplicity, we label the four squares of a superunit with numbers (see Figure S2 in the Supporting Information). The normal vectors are defined as $\bar{n}_1 = \frac{\bar{a}_1 \times \bar{a}_4}{a^2}$, $\bar{n}_2 = \frac{\bar{a}_1 \times \bar{a}_2}{a^2}$, $\bar{n}_3 = \frac{\bar{a}_1 \times \bar{a}_2}{a^2}$, and $\bar{n}_4 = \frac{\bar{a}_1 \times \bar{a}_4}{a^2}$, where $\bar{a}_i$ is the $i$-th edge vector of the square. Then, we can obtain the effective electric dipoles as $\bar{p}_1 = -p_1 \frac{\bar{n}_1 \times \bar{a}_1}{a}$, $\bar{p}_2 = -p_2 \frac{\bar{n}_2 \times \bar{a}_3}{a}$, $\bar{p}_3 = -p_3 \frac{\bar{n}_3 \times \bar{a}_3}{a}$, and $\bar{p}_4 = -p_4 \frac{\bar{n}_4 \times \bar{a}_1}{a}$. The effective magnetic dipoles are expressed as $\bar{m}_i = m_i \bar{n}_i$. The total effective electric and magnetic dipoles are $\bar{p}_{eff} = \sum \bar{p}_i = p_{eff,x}\hat{x} + p_{eff,y}\hat{y} + p_{eff,z}\hat{z}$ and $\bar{m}_{eff} = \sum \bar{m}_i = m_{eff,x}\hat{x} + m_{eff,y}\hat{y} + m_{eff,z}\hat{z}$, respectively. The surface current distributions for the 2D achiral metasurface and the L-handed enantiomers of the KCMMs are plotted in Figure 4(a-e). Before folding, the total effective electric dipole moment is along the $y$ direction, yet the total effective magnetic dipole moment is zero; hence, no chiroptical response occurs for the achiral metasurface. In the L-handed type-I KCMM, the induced magnetic dipoles are no longer antiparallel with each other, and hence, a net magnetic response occurs (Figure 4(c)). The total effective electric and magnetic dipoles are antiparallel to each other along the $y$-axis, contributing to the strong left-handed chiral response. For the L-handed type-II KCMM, the electric and magnetic dipoles experience out-of-plane rotations in the folding process. At the first resonant mode (6.78 GHz), the $y$ components of the effective electric and magnetic dipoles are antiparallel to each other, contributing to the strong left-handed chiral response. In contrast, the effective electric and magnetic dipoles are parallel at the second resonant mode (7.8 GHz), as shown in Figure 4(e), indicating the right-handed chiral response. Therefore, the type-II KCMMs flexibly switch their functionality between transparent and opaque states for circular polarized waves.

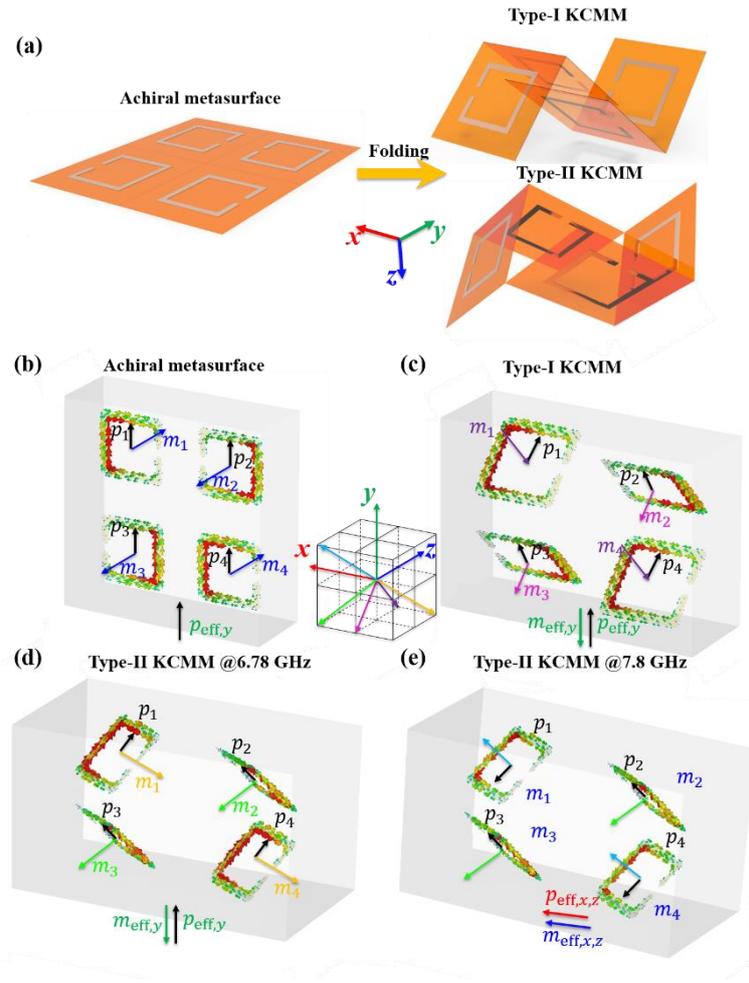

**Figure 4.** Surface current distributions and dipolar responses of the kirigami-based metamaterials. (a) The orientation of the four split-ring resonators (b) The surface current distributions at a resonant frequency of 6.9 GHz. Four resonators all working via the asymmetric modes with in-plane electric dipoles and out-of-plane magnetic dipoles. (c) The surface current distributions for the resonant modes of the L-handed type-I KCMM. The induced electric and magnetic dipoles have anti-parallel components, resulting in the left handedness. (d, e) The surface current distributions for the first (d) and second (e) resonant modes of the L-handed type-II KCMM. The induced electric and magnetic dipoles have anti-parallel components at 6.78 GHz but parallel components at 7.8 GHz, leading to the opposite handedness of the kirigami-based metamaterial at the two resonant frequencies.

To quantitatively analyze the nature of these chiral modes, we performed a multipole analysis by calculating the simulated surface current excited by the plane waves at normal incidence. Then, the multipolar expansions were calculated using the integral of the currents;[44-46] additional details can be found in the supplementary materials.

We performed a multipole decomposition to quantify their contributions. In our KCMMs, only the electric dipole **P**, magnetic dipole **M**, toroidal dipole **T**, and electric quadrupole $\mathbf{Q_e}$ provide significant contributions to the macroscopic electromagnetic responses, and all the other higher-order multipoles are relatively weak and can be neglected. The radiated powers of multipole components as a function of wavelength were calculated by integrating the current, and the results are plotted in Fig. 5. For the achiral metasurface (Fig. 5(a)), the y component of the electric dipole ($\mathbf{P_y}$) is the maximum component, and the other components show zero responses. The toroidal dipole $\mathbf{T_z}$ is weak, the electric quadrupole $\mathbf{Q_e}$ and magnetic dipole $\mathbf{M_y}$ are zero. When the achiral metasurface folds to the type-I KCMM (Fig. 5(b)), the electric dipole $\mathbf{P_y}$ and magnetic dipole $\mathbf{M_y}$ play dominant roles in comparison with the toroidal dipole $\mathbf{T_z}$, and the electric quadrupole $\mathbf{Q_e}$ is still extremely weak. The results agree with the qualitative analysis.

For the L-handed type-II KCMM, Fig 5(c, d) shows that the first resonant mode of type-II KCMM is similar to that of type-I KCMM, and the handedness switching in the second chiral resonant frequency is attributed to the appearance of other components of multipoles ($P_x$, $P_z$, $M_x$, $M_z$, $T_z$). Notably, the strong electric dipole component $P_z$, magnetic dipole component $M_z$ and toroidal dipole component $T_z$ do not directly contribute to the far-field radiation at normal incidence but may couple with the surface wave or other dark modes. This is the underlying mechanism of the second chiral resonance mode.

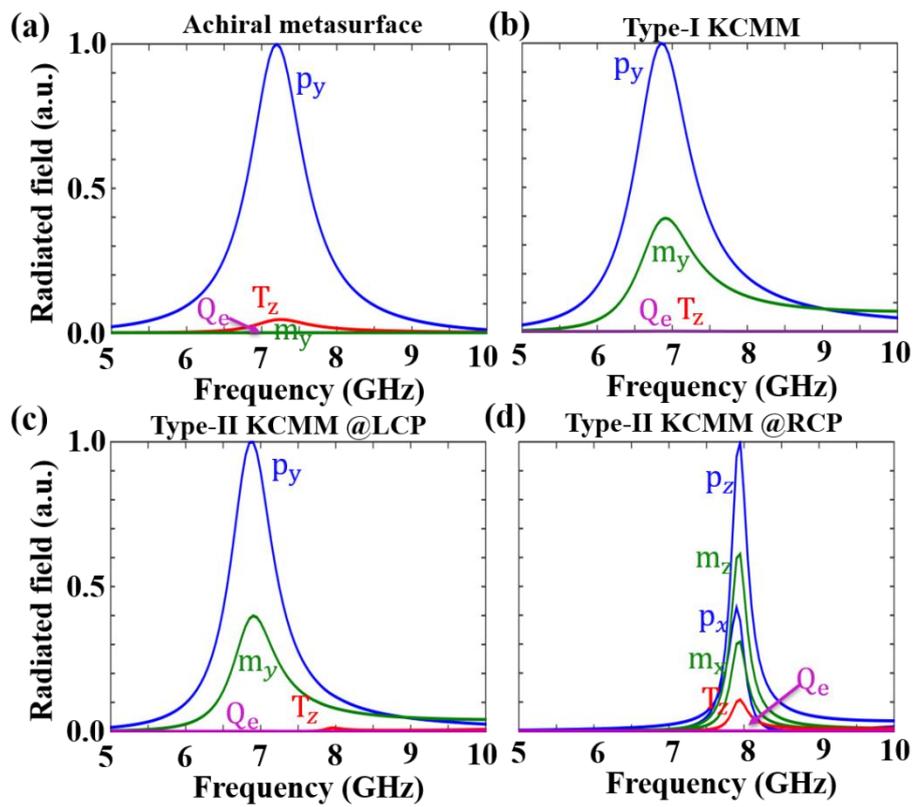

**Figure 5.** Amplitudes of the fields radiated by the four most dominant multipole components (electric dipole **P**, magnetic dipole **M**, toroidal dipole **T**, and electric quadrupole $Q_e$) that contribute to the polarization conversion response of the metamaterial. (a) The calculated scattering power of multipoles of the achiral metasurface. (b) The calculated scattering power of multipoles of the L-handed type-I KCMM. (c, d) The calculated scattering power of multipoles of the L-handed type-II

KCMM.

For the type-III KCMMs, the microscopic origin of the chiroptical responses is different from those for type-I and type-II. To understand the chiral performance, it is not sufficient to analyze the electric and magnetic dipoles only; toroidal multipolar analysis is needed[40,47]. The presence of the multipole pairs is visually detected in the distributions of the electric and magnetic fields inside the unit cell arising from circularly polarized excitation. At $f_1$ (7.12GHz), the field lines of the electric and magnetic fields are aligned parallel to the *y*-axis yet with opposite directions, which contributes to the electric and magnetic dipolar excitations with the net dipole moments collinearly oriented along the *y*-axis (Figure 6(a,b)). This is also the aforementioned microscopic origin of the chiroptical responses in type-I and type-II KCMMs. However, at $f_2$ (6.4GHz), the magnetic field is confined within a well-defined ring-like area where the field lines thread through the individual SRRs and form a closed loop (Figure 6(d)). Such a magnetic-field configuration is formed by poloidal currents flowing in the wire loops of the SRRs and is unique to the toroidal dipolar excitation with a net dipole moment aligned parallel to the *x*-axis. Compared with that in Figure 6(a), the distribution of the electric field shows a similar pattern but with field lines on the opposite sides of the SRR aligned antiparallel to each other, indicating an electric quadrupole excitation (Figure 6(c)). Therefore, the additional electric quadrupole and magnetic toroidal dipole excitations contribute to the broadband chiroptical performance of the type-III KCMMs. The analysis was performed using multipole decomposition for x and y polarization waves. Different multipoles contribute to the same resonant features in y polarization and x polarization incidence waves. As shown in Fig 6(e, f), the dominant multipoles are the electric and magnetic dipoles $\mathbf{P}_y$ and $\mathbf{M}_y$ at $f_1$. However, at $f_2$, the electric quadrupole $\mathbf{Q}_{\text{e-yz}}$ (electric quadrupole in the yz plane) and the toroidal dipole $\mathbf{T}_x$ show the largest moments, and those of the magnetic dipole $\mathbf{M}_y$ and electric dipole $\mathbf{P}_y$ are secondary. Circularly polarized light can be regarded as the decomposition of two linear polarizations. Hence, it is sufficient to perform multipole decomposition of the metamaterial's polarization conversion responses $t_{xy}$ and $t_{yx}$. From the multipole analysis shown in Figure 6, the electric dipole $\mathbf{P_y}$ and magnetic dipole $\mathbf{M_y}$ can be excited under $E_x$ and $E_y$ incident fields at $f_1$, respectively, indicating strong polarization conversion ($t_{yx}$ and $t_{xy}$). This implies that the circular dichroism that depends on *Im($t_{xy}$-$t_{yx}$)* is highly enhanced at this frequency. However, at $f_2$, the strong chiroptical response is attributed to the simultaneous excitation of $\mathbf{Q}_{\text{e-yz}}$ (electric quadrupole in the yz plane) and $\mathbf{T_x}$, from which the cross-polarization powers arises.

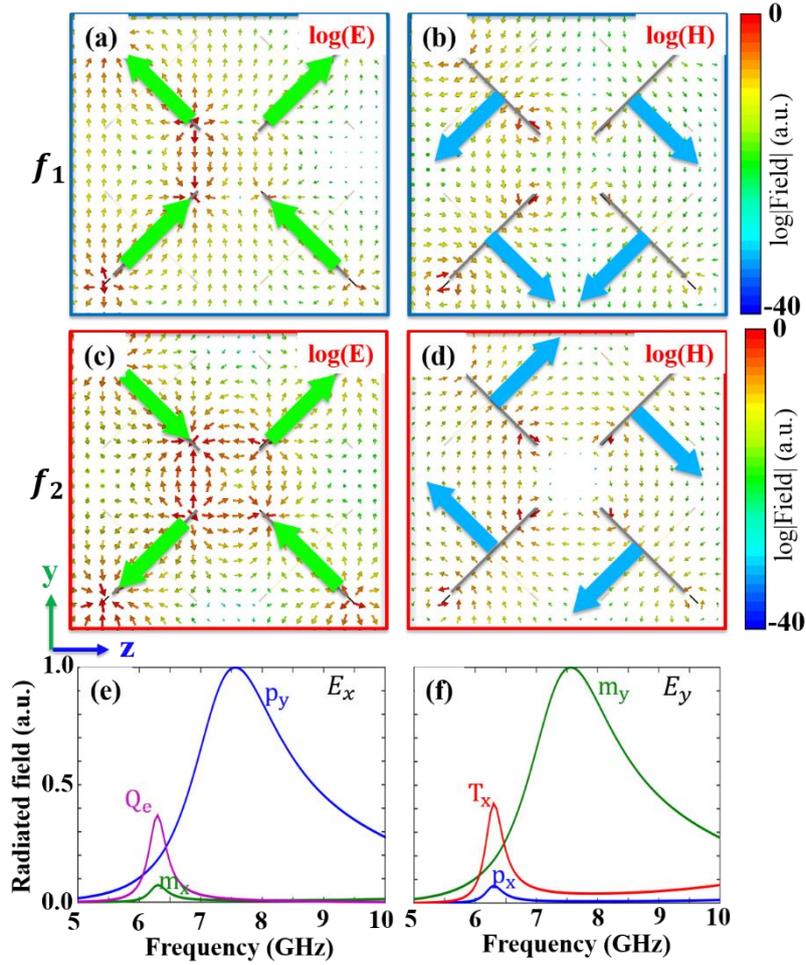

**Figure 6.** (a-d) Field distributions around a meta-atom of L-handed type-III KCMM when excited by LCP waves. (a, b) show the electric ($\log_{10}|E|$) (a) and magnetic ($\log_{10}|E|$) (b) field distributions at $f_1$ (7.12 GHz) (c, d) give the electric ($\log_{10}|E|$) (c) and magnetic ($\log_{10}|E|$) (d) field distributions at $f_2$ (6.4 GHz). The arrows indicate the directions of the fields. The green (blue) arrows show the orientation of the induced electric (magnetic) dipoles for the individual rings. (e, f) Amplitudes of the fields radiated by the four most dominant multipole components (electric dipole **P**, magnetic dipole **M**, toroidal dipole **T**, and electric quadrupole **Q**$_e$) that contribute to the polarization conversion response of the metamaterial under (e) x-polarized $E_x$ and (f) y-polarized $E_y$ incident light.

Toroidal moments have been found in nuclear and atomic physics and solid state physics. Recently, toroidal dipole excitations in metamaterials were observed, and the nonradiating feature of the toroidal geometry provides many interesting phenomena, with enhanced light-matter interaction and applications in lasers, ultrasensitive biosensors and nonlinear effects. We investigated the reconfigurable toroidal dipole moments. We also investigated the relationship between $\mathbf{T}_x$ and the folding angle $\theta$ by plotting the resonant strength of toroidal dipole excitation as a function of the folding angle. We also investigate the relationship between the toroidal dipole resonant frequency and the folding angle $\theta$. As shown in Figure 7(a), the resonant frequency of the toroidal dipole is sensitive to the folding angle. As the type III KCMM folds from a two-dimensional surface to 3D blocks, the resonant frequency increases to reach its maximum at approximately 45 degrees and then decreases. This feature is consistent with the simulated circular dichroism spectra in Figure 2(k). The radiated power of the toroidal dipole versus the folding angle is depicted in Figure 7(b). The radiated power of the toroidal dipole becomes relatively weak when the folding angle is too large or too small.

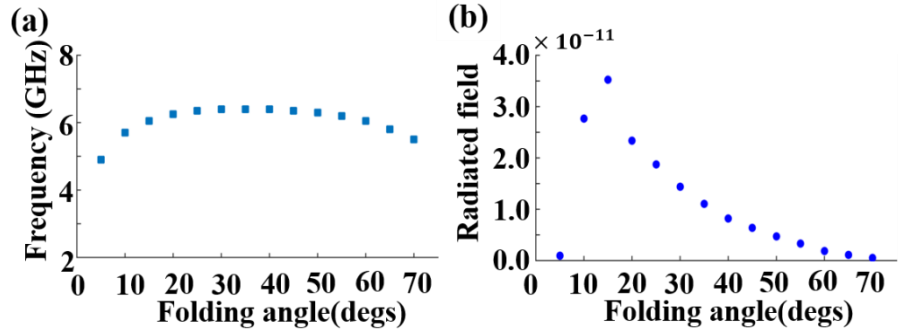

**Figure 7.** (a) Toroidal dipole $T_x$ resonant frequency as a function of folding angle. (b) Magnitude of the excited toroidal dipole component as a function of the folding angle at the instant of the electromagnetic resonance.

Moreover, due to the scaling properties of Maxwell's equations, the proposed strategy for reconfigurable electromagnetic performance can be extended to other frequencies, from millimeter-scale architectures in microwave regimes to tunable metasurfaces in terahertz regimes. We investigated one type of Micro-electro-mechanical-systems (MEMS) kirigami metamaterial. The right picture in Figure 8(a) depicts the basic structure of an SRR and two cantilever legs. The SRR can rotate as the cantilever legs bend in response to thermal changes. The Au/Cr (200 nm) layer is used for the SRRs, and the SiNx (400 nm) layer is used for the supporting layer and bending of the legs. The difference in the thermal expansion coefficients of SiNx and Au/Cr lead to bending of the cantilever supports. Rapid thermal annealing (RTA) at a specific temperature sets the orientation of the SRRs at a particular folding angle with respect to the substrate, as shown in Figure 8(a). In this way, it is possible to create a response over a large folding angle range from 0 to nearly 90°. CST MICROWAVE STUDIO was used to simulate the electromagnetic response. The conductivity of gold was set to 4.09e7 S/m, and the permittivity of SiNx was 7. To accurately simulate the bending of the bimaterial cantilevers, they were modeled as thin curved strips of constant length. The RTA at specific temperatures corresponds to particular folding angles [48-51]: $350\,°C$-$30°$, $400\,°C$-$40°$ $450\,°C$-$60°$, and $500\,°C$-$80°$.

Figure 8(b-d) show the results of simulations of the transmission spectra. No chiroptical response occurs for the achiral metasurface, as shown in Figure 8(b). Figure 8(c) shows transmission spectra corresponding to different temperatures: $350\,°C$, $450\,°C$, $450\,°C$, and $500\,°C$. The resonant frequencies gradually shift as the folding angle increases, indicating that the operating frequencies can be flexibly tuned by adjusting the deformation states. Figure 8(d) shows the corresponding circular dichroism (CD) spectrum.

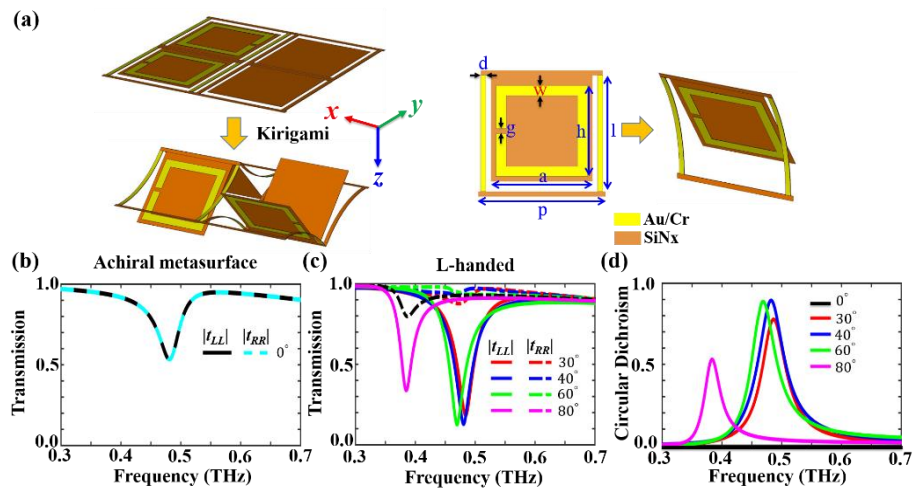

**Figure 8.** (a) Schematic view of a *MEMS* kirigami metamaterial structure and how the SRRs rotate as the cantilever legs bend. Parameters: p: 100 μm; length of the supporting SiNx plate a: 80 μm; SRR side length h: 72 μm; width of the bimaterial leg d: 4 μm; length of the bimaterial leg l: 92 μm; SRR line width w: 8 μm; and SRR gap

distance g: 4 μm. (b-c) Transmission spectra of the L-handed *MEMS* kirigami metamaterial. Black – 0 degrees, red – 30 degrees (350 °C), blue – 40 degrees (400 °C), green – 60 degrees (450 °C), and magenta – 80 degrees (500 °C). (d) Corresponding CD spectra for different folding angles.

**DISCUSSION**

We proposed and demonstrated a general class of kirigami-based chiral metamaterials whose electromagnetic functionalities can be switched between nonchiral and chiral states. By introducing cuts and selecting the connection points between neighboring meta-atoms, 2D achiral metasurfaces can be deformed to 3D shapes with significantly enhanced chiroptical responses. Highly efficient single-band, dual-band and broadband circular polarizers experimentally demonstrated switchable handedness realized by adjusting the deformation direction. The underlying mechanism was confirmed by detailed analyses of the excited electrical, magnetic and toroidal dipoles. Compared with the technique of origami, kirigami allows for the practitioner to exploit cuts in addition to folds to achieve large deformations and create complex 3D objects. With the ongoing development of micromanufacturing techniques, we expect our work to lead to alternate approaches to lightweight, reconfigurable, and deployable metadevices.

**CONFLICT OF INTEREST**

The authors declare no conflicts of interest.

**ACKNOWLEDGEMENTS**:

This work was sponsored by the National Natural Science Foundation of China under Grant No. 61625502, No. 61574127, No. 61601408, No. 61775193 and No. 11704332, the ZJNSF under Grant No. LY17F010008, the Top-Notch Young Talents Program of China, the Fundamental Research Funds for the Central Universities, and the Innovation Joint Research Center for Cyber-Physical-Society System.